\documentclass[aps,showpacs,twocolumn,prl]{revtex4}

\usepackage{amsmath}
\usepackage{amssymb}
\usepackage{graphicx}

\newcommand{\abs}[1]{\left|{#1}\right|}
\newcommand{\ket}[1]{\left|{#1}\right\rangle}

\newcommand{\braket}[2]{\langle{#1}|{#2}\rangle}
\newcommand{\ketbrad}[1]{\left|{#1}\rangle\langle{#1}\right|}
\newcommand{\Tr}{{\mathrm{Tr}}}
\newcommand{\nnn}{\bar{n}}

\newcommand{\ibid}{{\it ibid. }}

\begin{document}
\title{A computable measure of nonclassicality for light}

\author{J\'anos K. Asb\'oth$^{1,2}$, John Calsamiglia$^1$, Helmut Ritsch$^1$}

\affiliation{
$^1$ Institute of Theoretical Physics, University of Innsbruck, 
Technikerstrasse 25, A-6020 Innsbruck, Austria\\
$^2$ Research Institute of Solid State Physics and Optics,
Hungarian Academy of Sciences, 
H-1525 Budapest P.O. Box 49, Hungary
}

\date{\today}

\begin{abstract}
  We propose the \emph{entanglement potential} (EP) as a measure of
  nonclassicality for quantum states of a single-mode electromagnetic
  field. It is the amount of two-mode entanglement that can be
  generated from the field using linear optics, auxiliary classical
  states and ideal photodetectors. The EP detects nonclassicality, has
  a direct physical interpretation, and can be computed efficiently.
  These three properties together make it stand out from previously
  proposed nonclassicality measures. We derive closed expressions for
  the EP of important classes of states and analyze as an example the
  degradation of nonclassicality in lossy channels.
\end{abstract}


\maketitle

Quantum mechanics, and in particular quantum electrodynamics, enjoys
impeccable internal consistency and shows unmatched agreement with
experimental observations. Something remaining unresolved is
determining its borderline to the classical world, where quantum rules
are not observed.  A first step towards understanding the quantum to
classical transition is to find a description of classical states
within quantum theory. One can then ask how much
\emph{non-classicality} any given state possesses.

Coherent states \cite{glauber63} are generally accepted to be the most
classical states of optical fields. They reflect the wave-like nature
of classical light fields, are generated by classical currents, and
they are pointer states in realistic decohering environments.  Here we
will also adopt this notion and call a state \emph{classical} if it is
a coherent state, or a mixture thereof (see however
\cite{johansen04}). All other quantum states are referred to as
\emph{nonclassical}.
  
With recent progress in quantum optics it is becoming possible to
experimentally create an increasing wealth of light fields with
properties deviating more and more from coherent states
\cite{nonclass_experiment}. Such nonclassical states can be
distinguished from classical ones by many different features. Examples
are sub-Poissonian photon statistics, squeezing, photon number
oscillations or negative values of the Wigner function. Most of these
signatures can be quantified defining degrees of nonclassicality.

As a central problem, however, none of the above properties detects
nonclassicality infallibly \cite{prl89_283601}: e.g.~a so-called
Schr\"odinger cat state $\ket\alpha + \ket{-\alpha}$, with $\alpha\gg
1$ has Poissonian photon statistics, no squeezing, and yet is highly
nonclassical.  It is therefore desirable to have a general measure of
nonclassicality.  This could quantify a resource for a variety of
applications that are brought about by the quantum features of light
--- e.g.~high-precision interferometry, increased capacity of
communication channels, sub-wavelength lithography, and the plethora
of applications in communication and computation in Quantum
Information Theory (QIT).

A clear-cut ``universal'' approach to quantifying
nonclassicality has been formulated by Hillery \cite{pra35_725}. He
proposed the trace distance of a state $\sigma$ from the set of
classical states as a quantitative measure of its nonclassicality. As
desired it gives 0 for coherent-state mixtures and is nonzero for any
other state. However, Hillery's nonclassical distance is very hard to
compute, involving minimization over an infinite number of variables.
Consequently it has not yet been evaluated exactly for any
nonclassical state (for alternative `distance--based' measures and
approximations see \cite{approximations,marian_a,prl88_153601}).

Another universal measure, the nonclassical depth, was proposed by Lee
\cite{lee}. It is the amount of Gaussian smoothing required to
transform the Glauber P--function into a positive distribution and it
quantifies the noise necessary to wash out nonclassicality.
This measure, however, suffers from a lack of continuity: it lies
between $0$ and $1/2$ for Gaussian states, but is always equal to $1$
for any non-Gaussian pure state\cite{pra51_3340}, no matter how close
it is to being Gaussian.

In this Letter we propose a universal measure of nonclassicality, the
\emph{Entanglement Potential} (EP), that can be computed efficiently
for single-mode states of light. The measure is based on the
observation that coherent states are the only pure states that produce
uncorrelated outputs when split by a linear optical device
\cite{aharonov66,prerequisite}. We define the EP of a single-mode
state $\sigma$ as the amount of two-mode entanglement (or quantum
correlations) that can be produced from $\sigma$ and auxiliary
classical states with linear optics and ideal photodetectors. A
precise definition follows below.

Entanglement is a genuine quantum feature that gives rise to the most
striking ``nonclassical" effects \cite{schrodinger,bell}.  With the
advent of QIT entanglement has become a crucial resource for many
applications. The leading role of quantum optics in the study of the
foundations of quantum mechanics \cite{bell_experiment}
and in the implementation of many of the QIT protocols has triggered
in the recent years a lot of interest in the characterization,
generation, and distillation of entanglement between optical fields
\cite{modeEntanglement, prl90_047904}.  This provides the EP with a
direct physical meaning: single-mode nonclassicality as an
entanglement resource.  Moreover, it supplies a pool of results and
methods from QIT to be used in the study of classicality.

At first glance, the computation of EP seems prohibitively
complicated. For any nonclassical state $\sigma$ one has to find the
optimal linear optical transformation and auxiliary states
to create the most two-mode entanglement. However, as we show below,
the optimal linear optics entangler is the same for any state, and
consist of a single beamsplitter (BS) and an additional vacuum input.

\begin{figure}
\begin{center}
\includegraphics[width=9cm]{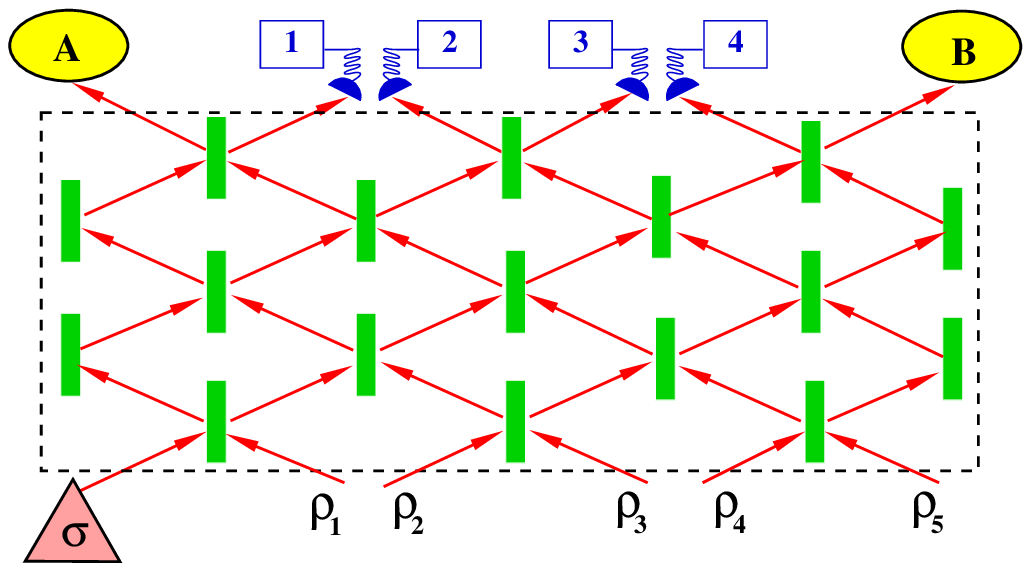}
\includegraphics[width=9cm]{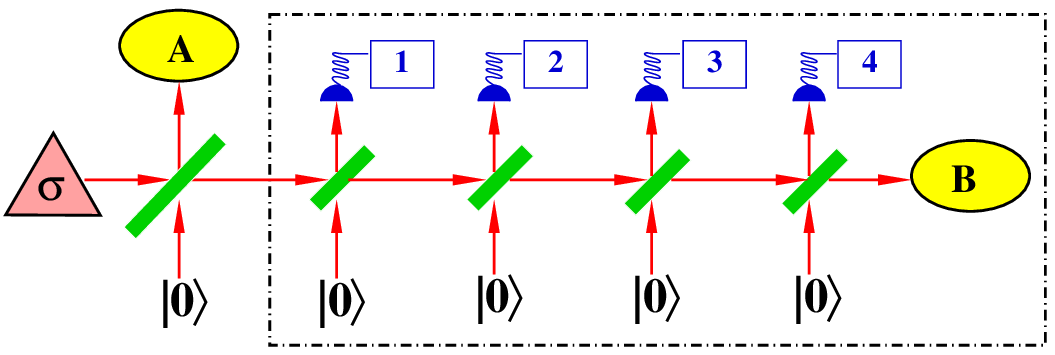}
\caption{ (top) A linear optics ``black box'' (dashed line) creating
  entanglement between A and B from a nonclassical input state
  $\sigma$.  Classical states enter via auxiliary input ports. Photon
  number measurements are made at the extra output ports.  (bottom) An
  equivalent form of the black box for empty auxiliary ports.  The
  region in the dash-dotted box is local to B.
\label{fig:Blackbox1} 
}
\end{center}
\end{figure}

A representation of the transformations that we allow in the
definition of the EP is shown in Fig.~\ref{fig:Blackbox1} (top).
A passive linear optical transformation can be modeled by a circuit of
BS's (including phase-shifters) \cite{prl73_58}. This transforms the
input state and auxiliary states according to a linear unitary map
$a_i'=\sum_{i} U_{ij} a_j$ between input and output mode annihilation
operators.  At the output two modes are sent to A(lice) and B(ob), all
other modes are measured by ideal photodetectors.  Note that nonlinear
optical devices and also linear operations conditioned on a
measurement outcome are excluded \cite{calsamiglia01}. Although such
linear feedback schemes cannot create nonclassicality --- they
preserve positivity of P-distributions---, we explicitly disallow them
since they can increase the amount of existing nonclassicality.

We first show that the auxiliary modes can be chosen to be in the
vacuum state. The displacement of one ancilla mode, i.e.
$\hat{D}(\alpha)\ket{0}=\ket{\alpha}$ with $\hat D(\alpha)=\exp(\alpha
a^\dagger - \alpha^* a)$, amounts to (local) displacement of all
output modes by amounts depending on the circuit of BS's.  Mixing the
displaced input modes translates to local mixing of the output
modes with additional classical communication. As these operations
cannot increase entanglement, vacuum ancillas are optimal for
entangling modes A and B.

Now, if the ancillas are vacuum, the circuit of BS's inside the box
can be simplified to a standard form. As shown in
Fig.~\ref{fig:Blackbox1} (bottom), this consists of a single BS that
splits the input mode in two modes corresponding to Alice and Bob, and
then a series of additional BS's further splits the signal into
various modes in Bob's side. All the measurements can be carried out
in Bob's auxiliary modes.  Local operations cannot increase
entanglement, and hence Bob can expect no advantage from splitting off
and measuring a part of the beam. The optimal entangling device is
therefore a single BS.  Although we currently lack of a general proof,
all examples that we checked analytically and numerically indicate
that the transmissivity of the optimal BS is $1/2$ independent of the
input state. We will denote by $U_{BS}$ the 50:50 BS transformation,
which induces the mapping $a=2^{-\frac{1}{2}} (a_A +a_B)$ on the input
mode annihilation operator.

Clearly, the EP is zero for classical states.  Moreover, any
decomposition of a given output two-mode mixed state in terms of pure
states $\rho=\sum_i p_i \ketbrad{\Psi_i}$ must be consistent
term-by-term with a pure input
$\ket{\psi_i}\ket{0}=U_{BS}^{-1}\ket{\Psi_i}$: this is necessary for
the corresponding input mixed state to have a vacuum auxiliary mode.
In particular, we find that any separable output state must correspond
to a convex combination of input states $\ket{\alpha_i}\ket{0}$,
i.e. to a classical input state. In other words, all nonclassical
input states, pure or mixed, will generate entanglement. This can be
seen as an extension of the results of \cite{prerequisite} that
nonclassicality at the input is a prerequisite to entanglement at the
output of a BS.  Since coherent displacement and phase shifting can be
realized on a single BS with an additional strong coherent beam, the
EP is invariant with respect to ``classical'' operations, as defined
in \cite{prl88_153601}.
 
To obtain a specific measure of nonclassicality an entanglement
measure has to be chosen. The value of EP of a state depends on this
measure, and different choices may give rise to different orderings
between states. Here we consider two alternatives.

A computable measure of nonclassicality for pure as well as for mixed
single-mode states is obtained by taking the logarithmic negativity
$E_\mathcal{N}$ \cite{pra65_032314} leading to the following
\emph{definition} for the \emph{Entanglement Potential},
\begin{equation}
EP(\sigma) \equiv E_\mathcal{N}(\rho_\sigma)
=\log_2||\rho_\sigma^{T_A} ||_1. 
\label{eq:EP}
\end{equation}
Here $\rho_\sigma=U_{BS} (\sigma\otimes\ketbrad{0})U_{BS}^\dagger$,
${\rho}^{T_A}$ denotes the partial transpose of $\rho$, and
$||\cdot||_1$ is the trace norm.  A nonzero value of $E_\mathcal{N}$
reveals that the state is nonseparable, however, the converse is not
true in general \cite{prl80_5239}. The so-called bound entangled
states are not detected by the partial transposition
criterion. Although examples of such entangled states exists for
two-mode non-Gaussian states \cite{boundE}, it remains an open
question whether bound entanglement can arise in our setup,
i.e.~whether or not $EP$ detects all nonclassical states.

An alternative entanglement measure that does detect all entangled
states is the relative entropy of entanglement $E_{RE}$
\cite{prl78_2275}.  We will call the induced nonclassicality measure
\emph{Entropic Entanglement Potential} ($EEP$), defined as
\begin{equation}
EEP(\sigma) \equiv \min_{\rho\in\mathcal{D}} \Tr\rho_\sigma
(\log_2\rho_\sigma-\log_2 \rho)
\label{EPP}
\end{equation}
where the minimization is carried out over the set $\mathcal{D}$ of
all two-mode separable states.  $EEP$ detects all nonclassical states,
and can be calculated for important classes of states. For finite
dimensional mixed states, it can be numerically computed by an
iterative procedure \cite{prl90_127904}.  For pure states the $EEP$
reduces to the von Neumann entropy generated in one of the output
arms.  Moreover, as we show in \cite{long_version}, $EEP$ gives a
lower bound to the nonclassical relative entropy 
distance \cite{marian_a}.

In the remainder of this Letter we calculate the $EP$ and $EEP$ for a
variety of nonclassical states.  A more comprehensive study and
comparison with other nonclassicality measures will be presented
elsewhere \cite{long_version}.

The $EP$ and the $EEP$ of a Fock state $\ket{n}$ are
\begin{align}
EP(n)  &= -n + 2  \log_2 \sum_{k=0}^n \sqrt{\binom{n}{k}},\\ 
EEP(n) &=  n - 2^{-n} \sum_{k=0}^n{\binom{n}{k} \log_2 \binom{n}{k}}. 
\end{align}
In the large-$n$ limit, the two diverge logarithmically and differ
only in an additive constant: $ EP(n) \approx \frac{1}{2}\log_2(2\pi
n)$;
$ EEP(n) \approx EP(n) - (1 - 1 / \ln 4) $.  
The entanglement potential can thus detect the nonclassicality of Fock
states, and shows increasing nonclassicality with increasing photon
number.

For any finite superposition of coherent states $\ket{\Psi}=\sum_k c_k
\ket{\alpha_k}$ both the $EP$ and the $EEP$ can be calculated exactly in
the nonorthogonal basis subtended by the $\ket{\alpha_k}$, by making use of
the ``metric tensor'' as in \cite{fortschr51_157}.  This yields complicated
formulas, a notable exception being the ``odd coherent state''
$\ket\alpha - \ket{-\alpha}$, for which
both the $EP$ and $EEP$ are 1, independent of $\alpha$.  If the coherent states 
are truly distinct $\abs{\alpha_i-\alpha_k}\gg 1$, the nonclassicality 
is determined by the probability amplitudes $c_i$, and the coherent 
amplitudes $\alpha_i$ bring only an exponentially small correction.
Such superpositions are often called Schr\"odinger cats (SC), 
and are typical examples of nonclassical states.
Neglecting the correction arising from the overlaps, 
we obtain for the $EP$ and the $EEP$ of a Schr\"odinger cat state:
\begin{align}
\label{eq:EP_pure_SC}
EP(\Psi) &= 2\log_2 \sum_k \abs{c_k} \\
EEP(\Psi) &= -\sum_k \abs{c_k}^2 \log_2 \abs{c_k}^2.
\end{align}
The most nonclassical superposition of $N$ coherent states
is that where the probability amplitudes are of equal magnitude:
in that case $EP= EEP = \log_2 N$.

A Gaussian state can be written as a displaced
squeezed thermal state: $\rho(\alpha,r,\phi,\nnn) = {\hat
D}(\alpha){\hat S}(r,\phi) \rho_{\nnn} {\hat S}(r,\phi)^\dagger {\hat
D}(\alpha)^\dagger.$ Here $\rho_{\nnn}$ is a thermal state of average
photon number $\nnn$, and $\hat S(r,\phi)=
\exp(\frac{1}{2}r[e^{i\phi}(a^\dagger)^2-e^{-i\phi}a^2])$ is the
squeezing operator.  It is well known \cite{pra47_4474} that Gaussian
states are classical for $r\le r_c$, where the nonclassicality
threshold is $r_c=\ln(2\nnn+1)/2$.  Using the results of Wolf et
al.\cite{prl90_047904}, it immediately follows that the $EP$ of a
general Gaussian state is given by the nonclassical part of the
squeezing,
\begin{equation}
\label{eq:gauss_EP}
EP( \rho(\alpha,r,\phi,\nnn)) = \frac{r-r_c}{\ln 2}.
\end{equation}
For Gaussian states $EP$ detects nonclassicality \cite{simon}, 
and it is a monotonous function of the nonclassical depth
\cite{lee}.

To calculate the $EEP$ we restrict ourselves to pure Gaussian states.
Displacement operators in the input map to local displacements at the
output, and similarly the squeezing maps to two-mode squeezing
followed by additional (local) squeezing in each mode. Hence, $EEP$
reduces to the single-mode entropy of the two-mode squeezed vacuum,
\begin{multline}
EEP(\rho(\alpha,r,\phi,0)) = \cosh^2 (r/2) \log_2 \cosh^2 (r/2)\\
 -\sinh^2(r/2) \log_2 \sinh^2(r/2).
\end{multline}
For strong $r\gg 1$  squeezing we again find approximate equality with 
the $EP$ up to a constant:  
$EEP(\rho_{\alpha,r,\phi,0})\approx  r/\ln 2 - (2-1/\ln 2)$.
For weak $r \ll 1$ squeezing, however, the $EEP$ increases quadratically    
with $r$: $EEP(\rho_{\alpha,r,\phi,0})\approx -r^2/2 \log_2(r/2)$.

A quantitative measure allows us to investigate 
how much nonclassicality is lost in a physical process. 
As an example, we study photon dissipation, which is the dominant
decoherence process for states propagating in an optical fiber.  Any
coherent state is decreased in amplitude by the factor
$\xi=\exp(-\gamma t)$, whereby the Glauber P function changes as
$P'(\alpha)= \xi^{-1}\, P(\xi^{-1/2}\alpha)$ \cite{Ulf}.

For a Schr\"odinger's cat undergoing photon loss the $EP$ can be
calculated exactly using the metric tensor \cite{fortschr51_157}.  For
weak dissipation, decoherence dominates power loss, and the
constituent coherent states $\ket{\xi \alpha_i}$ can still be
considered approximately orthogonal.  We then obtain
\begin{align}
\label{eq:neg_sc}
EP(t) \approx
\log_2\left(1+2\sum_{i<k}
\abs{\braket{\alpha_i}{\alpha_k}}^{1-\xi(t)} \abs{c_i}
\abs{c_k}\right).
\end{align}
Although the $EP$ or $EEP$ of SC states is independent of the
phase-space distance of the constituent coherent states, this distance
determines the decoherence behaviour.  For intermediate times
$\abs{\alpha_1-\alpha_2}^{-2}<\gamma t \ll 1$ the $EP$ of dissipated
SC is given by $EP(t) \approx \exp(-t/T_D ) 2\abs{c_1 c_2}/\ln
2$, with the well-known decoherence timescale $T_D =
2\abs{\alpha_1-\alpha_2}^{-2}\gamma^{-1}$.

The compact formula (\ref{eq:gauss_EP}) can be used to study
the nonclassicality of Gaussian states in Gaussian channels.
For example, for linear coupling to a heat bath of mean photon number
$n_T$ we find
\begin{equation}
EP(t) = -\frac{1}{2} \log_2\left[ e^{-\gamma t} e^{ -2 (r-r_c)} 
+ (1-e^{-\gamma t}) (2 n_T + 1) \right],\nonumber
\end{equation}
where $r$ is the initial squeezing, and $r_c$ is the initial
classicality threshold (as defined above).
We now concentrate on photon dissipation, i.e.~$n_T=0$.
For weak squeezing, $r-r_c \ll 1$, the above formula
reduces to an exponential decay,
$ EP(t) \approx \left(\ln 2\right)^{-1} (r-r_c) e^{-\gamma t}.$
For strong squeezing, $r-r_c \gg 1$, we find a different decoherence
behaviour.  Initially, $EP$ decreases linearly with time,
\begin{equation}
\gamma t \ll e^{-2(r-r_c)} \quad : \quad 
EP(t)\approx\frac{r-r_c}{\ln 2} -\frac{e^{2 (r-r_c)}}
{2 \ln 2}\gamma t.
\end{equation}
Note that the loss rate of $EP$ is exponentially large in the initial
squeezing.  After an exponentially short time $\tau= \gamma^{-1} e^{-2
(r-r_c)}$ the $EP$ of the initially highly squeezed Gaussian state
falls onto a general curve:
\begin{equation}
\gamma t > e^{-2(r-r_c)} \quad : \quad 
EP(t)\approx-\frac{1}{2} \log_2\left(1-e^{-\gamma t} \right).
\end{equation}
The initial squeezing now only adds a minor correction,
exponentially small in $r-r_c$, to the $EP$. On longer timescales,
$\gamma t \gg 1$, the $EP$ of strongly squeezed states also decreases
exponentially, $EP(t) = (2\ln 2)^{-1} \left(1-e^{-2(r-r_c)} \right)
e^{-\gamma t}.$ Here we explicitly included the small
correction in $r$, which is the only memory of the initial parameters.
\begin{figure}
\begin{center}
\includegraphics[angle=270,width=8cm]{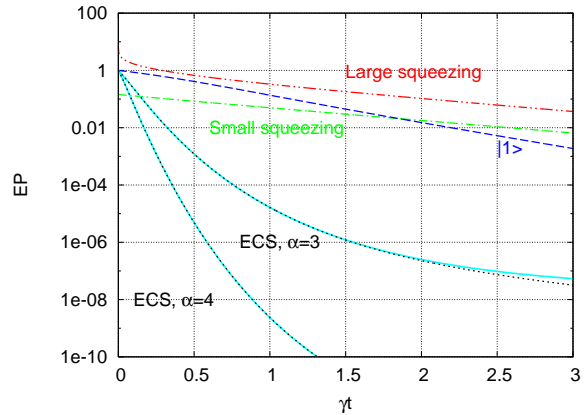}
\caption{ The $EP$ of Gaussian states with strong ($r-r_c = 5$,
  dash-dot-dot) or weak ($r-r_c = 0.1$, dash-dot) squeezing, the
  single-photon Fock state (dashed), and the even coherent states with
  $\alpha=3$ and $\alpha =4$.  For these latter states, besides the
  exact values (dots) the approximation of Eq.(\ref{eq:neg_sc})
  (continuous lines) is also shown.
\label{fig:decoh_color} 
}
\end{center}
\end{figure}

Figure~\ref{fig:decoh_color} shows the time dependence of $EP$ during
photon loss for various states.  Squeezed states retain their $EP$
better than other nonclassical states : although the weakly squeezed
state ($r-r_c=0.1$) initially has less $EP$ than either the
single-photon Fock state or the two examples of SC states, after $t >
2\gamma$ it is more valuable than these for entanglement
generation. The most fragile states are the SC's, which lose $EP$ on
short timescales given by the $T_D$ mentioned above.  We also notice
that (\ref{eq:neg_sc}) gives an excellent approximation of the
$EP$ for SC's.

Based on a broad and physically motivated definition we have
introduced a nonclassicality measure that grasps the essential feature
of single-mode nonclassical states: their potential to generate
entanglement by linear-optical means.  We have reduced the definition
to a simple and operational form, involving only a single
beamsplitter. The EP can be calculated analytically for a variety of
pure and mixed states, and efficient numerical methods exist for
general states. We have illustrated its use through a set of examples;
no other nonclassicality measure can be computed to cover {\it all} of
these. The definition of $EP$ in (\ref{eq:EP}) has brought up the
still open question of whether bound entanglement can be obtained from
single-mode nonclassical light, auxiliary classical states, linear
optics and photodetectors.  The study of the additivity properties of
the Entanglement Potential ---some examples show that it is
super-additive---, and the extension of the concept to multi-mode
nonclassical fields are subjects of further research.

\bibliographystyle{prsty}

\end{document}